\documentclass[12pt]{article}
\usepackage{epsfig}
\newcommand{\beq}{\begin{equation}}
\newcommand{\eeq}{\end{equation}}
\newcommand{\beqa}{\begin{eqnarray}}
\newcommand{\eeqa}{\end{eqnarray}}
\newcommand{\beqar}{\begin{eqnarray*}}
\newcommand{\eeqar}{\end{eqnarray*}}
\newcommand{\eg}{{\it e.g.,}\ }
\newcommand{\ie}{{\it i.e.,}\ }
\newcommand{\labell}[1]{\label{#1}}
\newcommand{\reef}[1]{(\ref{#1})}
\newcommand{\De}{\Delta}
\newcommand{\Si}{\Sigma}
\newcommand{\Dea}{\De+a^2\sin^2\theta}

\begin{document}
\addtolength{\baselineskip}{1.5mm} 

\thispagestyle{empty}

\hfill{}

\hfill{}

\hfill{CERN-TH/2001-122}

\hfill{hep-th/0105062}

\vspace{32pt}

\begin{center}
\textbf{\Large\bf Tubular Branes in Fluxbranes}\\

\vspace{48pt}

Roberto Emparan\footnote{roberto.emparan@cern.ch. Also at
Departamento de F{\'\i}sica Te\'orica, Universidad del Pa{\'\i}s Vasco,
Bilbao, Spain.}

\vspace{12pt}

\textit{
Theory Division, CERN, CH-1211 Geneva 23, Switzerland
}

\end{center}
\vspace{48pt}

\begin{abstract} 

We describe the construction of new configurations of self-gravitating
$p$-branes with worldvolume geometries of the form
$\mathbf{R}^{1,p-s}\times S^s$, with $1\leq s\leq p$, \ie
\textit{tubular branes}. Since such branes are typically unstable
against collapse of the sphere, they must be held in equilibrium by a
fluxbrane. We present solutions for string loops with non-singular
horizons, as well as M5-branes intersecting over such loops. We also
construct tubular branes which carry in their worldvolume a dissolved,
lower dimensional brane (as in the dielectric effect), or an F-string.
However, the connection between these solutions and related
configurations that have been studied earlier in the absence of brane
self-gravity, is unclear. It is argued that, at least in some
instances, the self-gravitating solutions do not appear to be able to
reproduce stable configurations of tubular branes.

\end{abstract}

\setcounter{footnote}{0}

\newpage

\section{Introduction}

One of the most important advances in string theory in the last years
is the realization that D-branes admit dual descriptions, in terms of
either open strings or closed strings. The open string description
lends itself to quantum field theoretic analysis, whereas the closed
string one finds its natural expression in terms of the different
classical supergravities. Such a duality first lead to a succesful
microscopic analysis of black holes \cite{strva}, and, from there, to
the celebrated AdS/CFT correspondence \cite{adscft}. Nevertheless, it
is probably fair to say that the closed string description lags, in
many respects, behind the open string one. 

To be more specific, consider the description of D-branes in terms of
Dirac-Born-Infeld actions, which derives from the requirement of
conformal invariance for open strings \cite{leigh}. The
Dirac-Born-Infeld picture allows one to easily consider some features
of the spacetime dynamics of D-branes, as well as aspects of brane
intersections and branes within branes. In some cases it provides a
handle on processes beyond the reach of string perturbation theory, as
for example, in the description of the decay of a RR field via the
nucleation of the spherical D-branes that couple minimally to such a
field \cite{teit}. It is also quite frequent that the dynamics of a
D-brane in a certain curved background can only be analyzed by means of
the ``test-brane'' approximation, where the gravitational backreaction
of the brane is neglected.

Clearly, the difficulties in the closed string description come from
the need to cope with a highly non-linear system, namely, general
relativity (usually in high dimensions) typically coupled to a variety
of scalar and $p$-form fields. Finding exact solutions to these field
equations is usually a very complicated task, but it is nevertheless an
imperative one, if the above mentioned duality is to be exploited in as
complete a way as possible.

Our purpose in this paper is to report on some progress in this
direction. We will present exact solutions to the equations of the
low-energy effective action of string/M-theory, which describe
$p$-branes whose worldvolume geometry is generically of the cylinder
form $\mathbf{R}^{1,p- s}\times S^s$, \ie \textit{tubular branes}. It
must be understood that the spherical part, $S^s$, does not wrap any
non-trivial cycle of the spacetime. As a consequence, given that branes
have a tension, their spherical part will tend to collapse down to a
point. Such a collapse can be prevented by the introduction of an
external field background, to which the brane couples. The simplest
analogy is that of an electron-positron dipole held in (unstable)
equilibrium by a carefully tuned electric field along the direction of
the dipole. In the presence of gravity, such background fields
concentrate under the influence of their self-gravity, and are referred
to as \textit{fluxbranes}. The best known instance among these is
provided by the Melvin universe \cite{melvin}. Analyses of fluxbranes
within the context of string theory include
\cite{gm,RT1,tsmelv,RT2,dggh,compo,chen,cogut,saffin,RT3}.

The study of gravitating spherical and tubular branes was initiated, in
a beautiful piece of work, in \cite{dggh}, which we review in the next
section. The context was that of Kaluza-Klein theories, or those that
could be related to them via dualities. However, several brane
configurations of interest are outside this range. For example, branes
whose horizons are regular (instead of null singularities), cannot be
constructed this way. Such branes are particularly important, since
their cores provide regular geometries that are amenable to a detailed
AdS/CFT description. The lowest dimensional instance, in the present
context, would be the dipole formed by two Reissner-Nordstrom black
holes with opposite charges, held in equilibrium by a background Melvin
field. The description of such a \textit{dihole}, including an
embedding as an intersection of branes and antibranes, has been given
in \cite{dihole,cet}. In one dimension higher, there exists a
five-dimensional string with a regular horizon: in this paper we will
present a new exact solution that describes this kind of string in the
shape of a loop (section \ref{regloop}). We also construct a
configuration where three M5-branes (with different charges) intersect
over one such loop (section \ref{threem}). Unfortunately, the
presumably most interesting configuration, a regular six-dimensional
string loop, is beyond the approach herewith followed.

In a slightly different direction, given that D-branes can carry along
their worldvolume field excitations that correspond to lower branes, or
strings, dissolved in them (branes within branes), one can ask whether
it might be possible to obtain solutions that describe tubular, or
spherical, branes with a net charge\footnote{Note that the spherical or
tubular branes themselves, as the analogues of dipoles, do not carry a
net charge.}. This net charge would correspond to a lower dimensional
brane, which, under the presence of the fluxbrane, has blown up into
the tube, and dissolved itself in it. Such configurations were first
analyzed, using the Dirac-Born-Infeld description, in \cite{stritun}.
More recently, they have been extensively studied within the context of
the ``dielectric effect'' for D-branes \cite{robs}. Another system
recently studied, where a tubular brane (called a ``supertube''
\cite{stubes}) can support itself against collapse without the need of
an external field, will also be of interest to us here. In section
\ref{blow} we will present the construction of a number of supergravity
solutions for D-brane tubes with this sort of charges. However, the
analysis raises some doubts about whether these supergravity solutions
adequately describe the supertubes of \cite{stubes}, or even the
dielectric branes of \cite{robs}. It is unclear whether these
self-gravitating solutions can describe tubular branes other than those
in unstable equilibrium.

A remark on notation is in order. Spacetimes of different
dimensionalities will abound, and a number of functions will be used,
some of which will depend on the dimensionality of the space one is in.
Then, when we employ the function defined below as $\Delta=r^2-a^2-\mu
r^{n}$, we will be careful to specify which value of $n$ is adequate in
each situation. On the other hand, the function
$\Si=r^2-a^2\cos^2\theta$, will always be defined as this, whatever the
dimension of the spacetime.

Finally, we note that when this work was in its last stages, a paper
appeared in which spherical test branes in the presence of fluxbranes
were studied \cite{strgut}. Even more recently, another paper has
appeared, which has more significant overlap with our work here
\cite{chc}. In particular, the construction of the D4-brane blown up
into a D6 in \cite{chc} is equivalent to our construction of the
D0-brane blown up into a D2 sphere, in section \ref{blow} .

\section{Spherical and Tubular Branes}\label{prewash}

In this section we describe, following \cite{dggh}, the construction of
$p$-branes with worldvolume geometry ${\bf R}^{1,p-s}\times S^s$, in a
spacetime of $p+4$ dimensions. We will also provide the explicit metrics and
fields for the solutions. We shall begin with the case $p=s$, \ie
spherical branes.

\subsection{Spherical Branes}

The starting point is the Euclidean continuation of the rotating black
hole in $D=p+4$ dimensions, with rotation in only one plane \cite{mype}.
To this, a flat time direction is added, resulting in a metric 
\beqa
ds^2_{D+1}&=&-dt^2+{\Dea\over \Si}\left(dx-{\mu a\sin^2\theta r^{5-
D}\over\Dea}d\varphi\right)^2
+\Si\left({dr^2\over\De}+d\theta^2\right)\nonumber\\&+&
{\Si\De\sin^2\theta\over \Dea} d\varphi^2+r^2\cos^2\theta
d\Omega^2_{D-4} \,,
\labell{start}
\eeqa where 
\beqa 
\De&=&r^2-a^2-\mu
r^{5-D}\,,\nonumber\\ \Si&=&r^2-a^2\cos^2\theta \,.
\eeqa 
For
$D\geq 5$, which will be the case of interest here, the polar 
coordinate $\theta$ runs along $0\leq\theta\leq \pi/2$, while
$\varphi\in[0,2\pi]$ is an azimuthal angle. On the other hand, given
the presence of a bolt (a ``Euclidean horizon'') at the largest root of
$\Delta$, $r=r_+$, then the coordinate $r$ takes only values $r\geq
r_+$, and the coordinate $x$ must be identified periodically, with a
period equal to the inverse of the black hole temperature. Further
details on the geometric aspects of these identifications can be found
in\cite{dggh}. Here we directly proceed to dimensionally reduce this
solution along the direction of $x$. This  results in a $D$-dimensional
solution of a theory with action\footnote{Our conventions differ from
\cite{dggh} in the sign and normalization of the dilaton.}
\beq 
I={1\over 16\pi}\int d^Dx\;\left[R-{1\over 2}(\partial\phi)^2-e^{
{\sqrt{2(D-1)\over D-2}}\phi}F^2\right] \,.
\eeq 
The Einstein-frame
metric of this solution is 
\beqa 
ds^2_D&=&\left({\Dea\over \Si}\right)^{1\over D-2}\left[-
dt^2+\Si\left({dr^2\over\De}+d\theta^2\right)+ r^2\cos^2\theta
d\Omega^2_{D-4}\right]\nonumber\\ &+&\left({\Si\over \Dea}\right)^{D-
3\over D-2}\De\sin^2\theta d\varphi^2 \,,
\labell{branesph}
\eeqa 
while the
Kaluza-Klein gauge potential and dilaton are 
\beqa 
A_{\varphi}&=&-{\mu
a\sin^2\theta \over 2r^{D-5}(\Dea)} \,,
\labell{kkpot}\eeqa 
and 
\beq
e^{\phi}=\left({\Dea\over\Si}\right)^{\sqrt{D-1\over 2(D-2)}} \,.
\labell{KKdil}
\eeq 

A dual theory, where magnetic monopoles are mapped into $(D-4)$-branes
that act as electrical sources for a $(D-2)$-form field strength, is
obtained by tranforming
\beq
\tilde\phi=-\phi,\qquad F_{[D-2]}=2e^{
{\sqrt{2(D-1)\over D-2}}\phi}\ast F\,,
\eeq
the metric remaining invariant. The dual theory is
\beq
I={1\over 16\pi}\int d^5x\;\left(R-{1\over 2}(\partial\tilde
\phi)^2-{1\over
2(D-2)!}e^{
{\sqrt{2(D-1)\over D-2}}\tilde\phi}F_{[D-2]}^2\right)\,.
\eeq

Then, the electric dual to \reef{kkpot} is
\beq
A_{t\psi_1\dots\psi_{D-4}}=
{\mu a\cos^{D-3}\theta\;\epsilon(\Omega_{D-4})\over \Si}\,,
\eeq
where $\epsilon(\Omega_{D-4})$ is the volume factor for the 
$(D-4)$-sphere, parametrized by angular coordinates $\psi_i$. 
For this theory, the Kaluza-Klein interpretation is lost.

The reasoning used in \cite{dggh} to argue that the solution
\reef{branesph}, \reef{kkpot}, \reef{KKdil}, describes a spherical
distribution of monopole charges was based on the topological structure
of the higher dimensional solution. Here we follow a different,
complementary approach, that works directly with the reduced solution.
It is perhaps less elegant, but it can be applied as well to solutions
that do not admit a Kaluza-Klein interpretation (\eg for other values
of the dilaton coupling). For the case of $D=4$, this approach has been
used in \cite{senD6} and \cite{dihole}, where a closely related
detailed analysis can be found. The idea is to explicitly exhibit that
on a sphere $S^{D-4}$, specified by $r=r_+$, with $r_+$ being the
largest root of $\De$, and by $\theta=0$, there lie magnetic monopole
charges. The charges are uniformly distributed over the sphere.

In order to study the solution near this sphere
it is convenient to first perform the change of coordinates
\beqa
r&=&r_++{\rho\over 2}(1+\cos\bar\theta)\nonumber\\
\sin^2\theta&=&{2\rho\over\Delta'_+}(1-\cos\bar\theta)\,,
\labell{transform}\eeqa
where 
\beq
\Delta'_+\equiv {d\Delta\over dr}|_{r=r_+}=2r_++(D-5)\mu
r_+^{4-D} \,.
\eeq
Then one takes $\rho$
to be much smaller than any other length scale involved, so as to get
close to
the locus ($r=r_+,\theta=0$). In this way, one finds that the metric in the
region close
to this sphere approaches
\beqa
ds^2_D&=&g(\bar\theta)^{1\over D-2}\left[\left({\rho\over
q}\right)^{1\over D-2}(-dt^2+r_+^2d\Omega^2_{D-4})+
\left({\rho\over q}\right)^{-{D-3\over D-
2}}(d\rho^2+\rho^2d\bar\theta^2)\right]\nonumber\\
&+&g(\bar\theta)^{-{D-3\over D-
2}}\left({\rho \over q}\right)^{-{D-3\over D-
2}}\rho^2\sin^2\bar\theta d\varphi^2\,,
\labell{nearmet}\eeqa
where we have defined $q\equiv {r_+^2-a^2\over\Delta'_+}$. The dilaton
and gauge potential approach
\beqa
e^\phi&=&\left({\rho g(\bar\theta)\over q}\right)^{\sqrt{D-1
\over 2(D-2)}},\nonumber\\
A_\varphi&=&-{a \over \Delta'_+}
{q(1-\cos\bar\theta)\over g(\bar\theta)}\,.
\labell{nearpot}
\eeqa
If the factor $g(\theta)$ were replaced by $1$, this would be precisely
the solution near the core of a set of Kaluza-Klein magnetic monopoles,
uniformly distributed over the sphere $S^{D-4}$ \footnote{Since $\rho\ll
r_+$, near the core this sphere looks very large, and nearly flat.
Effectively, $r_+^2d\Omega^2_{D-4}\to\sum_{i=1}^{D-4}dx_i^2$.}. The
configuration,
though, is angularly distorted by the presence of
\beq
g(\bar\theta)={1\over 2}\left[1+\cos\bar\theta+\left({2a\over
\Delta'_+}\right)^2(1-
\cos\bar\theta)\right]\,.
\labell{gtheta}\eeq 

This distortion is also related to another feature of the solution
\reef{branesph}, namely, the presence of conical singularities.
To see these, observe that $r=r_+$ is a fixed-point set of the generator
of rotations $\partial_\varphi$, hence a part of its axis. The
conical singularity is then evidenced by the fact that the ratio of
circumference to proper radius,
\beq
\lim_{r\to r_+}{2\pi\over\sqrt{g_{rr}}}
{d\sqrt{g_{\varphi\varphi}}\over dr}={\pi\Delta'_+\over
a}\,,
\labell{cone}\eeq
does not equal the canonical value $2\pi$ if $\mu\neq 0$. As a matter
of fact, such conical singularities are typical of situations where
there are unbalanced forces. In the present case, this is indeed
expected, since the monopoles attract each other, both by magnetic and
gravitational forces\footnote{Bear in mind that monopoles at antipodal
points on the sphere have opposite charges, since for $D>4$ the sign of
the magnetic charge for a two-form field strength depends on a choice
of orientation \cite{dggh}. This point is perhaps more obvious for the
electric brane solutions.}. 

The sphere $S^{D-4}$ and its interior lie at $r=r_+$. In the interior,
$\theta$, or more precisely, $\cos\theta$, plays the role of the radial
coordinate inside the sphere: notice that, at fixed $r=r_+$, the
``volume radius'' of the sphere is proportional to $\cos\theta$. In
particular, the center of the sphere is at $\theta=\pi/2$, while the
surface, where the monopoles lie, is at $\theta=0$. The conical
singularity extends over all of the interior, but apart from this, the
interior of the sphere presents no other singularities.

Let us also remark that, if instead of considering the limit of small
$\rho$ above, we had sent $a\to\infty$, while keeping $r-r_+$,
$a\sin^2\theta$, and $q$ finite, then we would have recovered precisely
the solution for a planar (infinite) distribution of magnetic monopoles.
Hence we see that the parameter $a$ controls, through $r_+$, the size of
the monopole sphere.

In order to balance the system, and cancel the conical singularities,
one can introduce an external magnetic field. The way to introduce a
magnetic field fluxbrane in a KK setup involves twisting the
compactification direction \cite{dgkt}. Effectively, one shifts the
coordinate $\varphi\to\varphi-Bx$, and reidentifies points
appropriately. Then, upon reduction, the solution becomes
\beq
ds^2_D=\Lambda^{1\over
D-2}\left[-
dt^2+\Si\left({dr^2\over\De}+d\theta^2\right)+
r^2\cos^2\theta d\Omega^2_{D-4}\right]+
{\De\sin^2\theta\over \Lambda^{D-
3\over D-2}} d\varphi^2\,,
\eeq
where
\beq
\Lambda={\Dea+ 2B a\mu r^{5-D}\sin^2\theta+B^2\sin^2\theta
[(r^2-a^2)^2+\De a^2\sin^2\theta]
\over \Si}\,.
\eeq
The gauge potential is now
\beqa
A_{\varphi}&=&-{\mu a r^{5-D} +B[(r^2-a^2)^2+\De
a^2\sin^2\theta]\over 2 \Lambda \Si}\sin^2\theta\,,
\labell{abpot}\eeqa
and $e^{\phi}=\Lambda^{\sqrt{D-1\over 2(D-2)}}$. 

Let us now see whether we can tune the field $B$ in such a way that the
conical singularities are removed. With the $B$ field on, the
circumference/radius ratio becomes, as one approaches the portion of
the axis along $r=r_+$,
\beq
\lim_{r\to r_+}{2\pi\over\sqrt{g_{rr}}}
{d\sqrt{g_{\varphi\varphi}}\over dr}={\pi\Delta'_+\over
a+B(r_+^2-a^2)}\,.
\eeq
Hence, the axis will be regular if we choose
\beq
B={\Delta'_+-2a\over 2(r_+^2-a^2)}={(D-3)r_++(D-5)a\over 2
r_+(r_++a)}\,.
\labell{chooseb}\eeq

Analyze now the solution in the region near ($r=r_+,\theta=0$), by means
of the same transformation \reef{transform}. One finds the metric
becomes exactly like \reef{nearmet}, but now with 
\beq
g(\bar\theta)={1\over 2}\left[1+\cos\bar\theta+4
\left({a+B(r_+^2-a^2)\over
\Delta'_+}\right)^2(1-
\cos\bar\theta)\right]\,.
\eeq
Therefore, if the field is tuned to the value \reef{chooseb}, we find
$g(\bar\theta)=1$, and the angular distortion of the core disappears
completely.
The solution approaches exactly the core of the
monopole. On the other hand, the gauge potential near the core,
\beq
A_\varphi=-{a+B(r_+^2-a^2)\over \Delta'_+ g(\bar\theta)}q(1-
\cos\bar\theta)\,,
\eeq
recovers its precise spherical monopolar form for the equilibrium value
for $B$, \reef{chooseb}.

The way we have presented our results will be the one we shall follow
throughout the paper. Namely, in order to show that a configuration
describes a spherical brane, it suffices to consider first the simpler
situation where the background fluxbrane is absent. The solution near
the core is slightly distorted, and presents a conical singularity, but
nevertheless it already possesses the main features of the spherical
brane. Having identified the configuration properly, it is a
straightforward matter to balance the system by immersing it in the
field of a fluxbrane.

\subsection{Tubular Branes}

This construction can be modified very easily to construct tubular
$p$-branes with worldvolume geometry ${\bf R}^{1,p-s}\times S^s$. To
this effect, since we are considering an $s$-dimensional sphere, first
set $D=s+4$ in the metric \reef{start}. Then, simply add a number $p-s$
of flat dimensions along with the time coordinate. Reduction along the
$x$ coordinate results into the required tubular $p$-brane. The
solution lives in a spacetime of $p+4$ dimensions. 

In particular, for ten-dimensional string theory, one obtains in this
way a whole range of tubular D6-branes, with worldvolume geometry ${\bf
R}^{1,6-s}\times S^s$. The case of $s=0$ is the D6-$\bar{\mathrm{D}6}$
state of \cite{senD6}. T-duality can be applied along some or all of
the $6-s$ flat worldvolume directions, and results into delocalized
(smeared) configurations of other tubular Dp-branes, with $p\leq 6$. In
conjunction with $S$-duality, one can as well construct delocalized
configurations for fundamental strings (as noted in \cite{dggh}) and
solitonic fivebranes. Some particular instances of this construction
can be found in \cite{janmuk}.

\section{Other dilaton couplings, and non-singular string
loops}\label{regloop}

There are several shortcomings to the construction described in the
previous section. Notice that, if the brane is to be a localized one,
then the dimension of the spacetime it lives in is restricted to the
value $p+4$. It appears to be very difficult to overcome this
restriction, and in fact we will have nothing to add in this respect.
Another constraint arises on the allowed values of the dilaton
coupling. These are restricted to those available by Kaluza-Klein
reduction. In particular this means that the above solutions cannot be
used to describe large classes of intersecting branes. Among the brane
intersections, a particularly interesting class is constituted by those
that intersect over a regular horizon (even though each individual,
delocalized brane may have a null singular core). Examples include the
Reissner-Nordstrom black hole in four dimensions, which can be embedded
as, \eg an intersection of four D3-branes, and the five-dimensional and
six-dimensional black strings, which appear at a triple intersection of
M5-branes \cite{Ts}. Now, the question is whether one can construct
configurations where the branes intersect over a loop or a sphere. The
lowest dimensional example is that where, instead of a loop or a
sphere, the branes intersect over a dipole formed by two black holes in
four dimensions, with equal masses and opposite charges---a dihole. The
dihole solution formed by the intersection of two sets of four
D3-branes each, has been already constructed in \cite{cet}. Going to a
one-dimensional intersection, the triple intersection of M5-branes over
a string loop will be built in the next section. Here, as a first step,
we build string loops in five dimensions with arbitrary dilaton
coupling $\alpha$. In the particular case of $\alpha=0$, where the
dilaton decouples, the string loop will have a non-singular horizon. It would also be interesting to build a six-dimensional
loop of string with a regular horizon. However, this requires a
solution outside the class of spherical $p$-branes in $p+4$ dimensions
(instead, $p+5$ would be needed), and therefore it is beyond the
techniques employed here.

Hence, let us consider the theories and solutions in the previous
section, for the case $D=5$. They are a particular case of dilatonic
theories
\beq
I={1\over 16\pi}\int d^5x\;\left(R-{1\over
2}(\partial\phi)^2-e^{\alpha\phi}F^2\right)\,,
\labell{dil5d}\eeq
namely, the ones with $\alpha=2\sqrt{2/3}$. For a given value of
$\alpha$, it is convenient to define a parameter $N$ via
\beq
\alpha^2={4\over N}-{4\over 3}\,.
\eeq
When the solutions are realized as brane intersections, $N$ is actually
the number of intersecting branes. The case of Kaluza-Klein theory
corresponds to $N=1$, and the non-dilatonic theory is obtained for
$N=3$. The dual theory, where magnetic monopoles are mapped into
electric strings, is obtained by changing to $\tilde\phi=\phi$ and
$H=2e^{\alpha\phi}\ast F$,
\beq
I={1\over 16\pi}\int d^5x\;\left(R-{1\over 2}(\partial\phi)^2-{1\over
12}e^{\alpha\tilde\phi}H^2\right)\,.
\eeq

We have found a solution to these theories for arbitrary $\alpha$, \ie
arbitrary $N\leq 3$, which
generalizes the Kaluza-Klein loop of monopoles that was
described in the previous section. The metric is
\beqa
ds^2&=&\left({\Dea\over \Si}\right)^{N/3}\left[-
dt^2+
r^2\cos^2\theta d\psi^2
+{\Si^N\over
[\De+(\mu+a^2)\sin^2\theta]^{N-1}}
\left({dr^2\over\De}+d\theta^2\right)\right]\nonumber\\
&+&\left({\Si\over \Dea}\right)^{2N/3}\De\sin^2\theta d\varphi^2\,,
\eeqa
and the dilaton and magnetic field
\beqa
e^{\phi}&=&\left({\Dea\over \Si}\right)^{N\alpha/2}\,\nonumber\\
A_\varphi&=&-{\sqrt{N}\over 2}{\mu a\sin^2\theta \over \Dea}\,.
\eeqa
The electric dual strings have
2-form potential $(H=dB)$
\beq
B_{t\psi}=-\sqrt{N}{\mu a\cos^2\theta\over\Si}\,.
\eeq
In this case, $\Delta=r^2-a^2-\mu$, hence $r_+=\sqrt{\mu+a^2}$. Most
aspects of this solution involve only a slight change from the
Kaluza-Klein solution $N=1$, except for one: the metric coefficients
$g_{rr}$ and $g_{\theta\theta}$ have acquired a new term, that is
absent for $N=1$. Actually, this new factor is analogous to the one
that appears in the
four dimensional diholes \cite{bonnor,dg,dihole}. 

In the same fashion as described in the previous section, one sees that
the string loop is located at $r=r_+=\sqrt{\mu+a^2}$ and $\theta=0$.
The interior of the loop is the two-dimensional disk at $r=r_+$. On it,
one moves from the boundary to the center by varying
$\theta\in[0,\pi/2]$, and in the angular direction by varying
$\psi\in[0,2\pi]$. The transformation \reef{transform}, in the limit of
small $\rho$, reproduces the (distorted) geometry near the core of
five-dimensional magnetic monopoles distributed along a circle, or
alternatively, a five-dimensional electric string loop.

Again, these configurations present the expected conical singularities, which
will be removed by the addition of a fluxbrane background. The
Kaluza-Klein procedure of twisting the direction of reduction cannot be
applied to these cases. However, a solution-generating transformation
can be found for the theories \reef{dil5d}, that extends to five
dimensions the Harrison transformations of \cite{dgkt}.
Given an axisymmetric solution to \reef{dil5d}, where the only non-zero
component of the gauge field is $A_\varphi$, and the metric $g_{ij},
g_{\varphi\varphi}$, $(i,j\neq \varphi)$, a new solution is generated by
\beqa
g'_{ij}&=&\lambda^{N/3}g_{ij}\,\qquad 
g'_{\varphi\varphi}=\lambda^{-2N/3}g_{\varphi\varphi}\,\nonumber\\
e^{\phi'}&=&e^\phi \lambda^{N\alpha/2}\,\qquad
A'_\varphi={N\over 2 B \lambda}\left(1+{2\over N}B A_\varphi\right)+
k\,,
\eeqa
where
\beq
\lambda=\left(1+{2\over N}BA_\varphi\right)^2+{1\over
N}B^2g_{\varphi\varphi}e^{-\alpha\phi}\,,
\eeq
and $k$ is a constant that can be adjusted to fix Dirac strings: we
will set $k=-N/(2B)$. Such a transformation generates a magnetic
fluxbrane background, with the field strength along the axis
asymptotically given by the parameter $B$. By dualization, one obtains
an electric fluxbrane background associated to the three-form $H$.

It is straightforward to apply this transformation to the monopole
string loops \reef{monostring}, and then dualize to obtain an electric
string loop. Doing so, we obtain
\beqa
ds^2&=&\Lambda^{N/3}\left[-
dt^2+
r^2\cos^2\theta d\psi^2
+{\Si^N\over
[\De+(\mu+a^2)\sin^2\theta]^{N-1}}
\left({dr^2\over\De}+d\theta^2\right)\right]\nonumber\\
&+&\Lambda^{-2N/3}\De\sin^2\theta d\varphi^2\,,
\labell{monostring}\eeqa
and $e^\phi=\Lambda^{N\alpha/2}$, with
\beq
\Lambda={\Dea+{2\over\sqrt{N}}B\mu a\sin^2\theta+{1\over
N}B^2\sin^2\theta[(r^2-a^2)^2+\De a^2\sin^2\theta]\over\Si}\,,
\eeq
and
\beq
B_{t\psi}=-\sqrt{N}\left[{\mu a\cos^2\theta\over\Si}\left(1-
{aB\over\sqrt{N}}\sin^2\theta\right)
\left(1+(1-{2/\sqrt{N}})
{aB\over\sqrt{N}}\sin^2\theta\right)-{1\over
N}Br^2\cos^2\theta\right]\,.
\eeq
In particular, the last term in the potential corresponds to the
asymptotic electric fluxbrane.

At this point, we can tune the fluxbrane strength $B$ so as to balance
the system and cancel the conical singularities on $r=r_+$. The field
that achieves this is
\beq
B=\sqrt{N}{\sqrt{\mu+a^2}-a\over \mu}\,.
\eeq
A choice of sign for $\sqrt{\mu+a^2}$ has been made here. As in
\cite{dihole}, we have chosen the one that yields $B\to 0$ as
$a\to\infty$. 

For this value of the field, the solution near the loop becomes
precisely
\beq
ds^2=\left({\rho\over q}\right)^{N/3}
(-dt^2+r_+^2d\psi^2)+\left({\rho\over
q}\right)^{-2N/3}[d\rho^2+\rho^2(d\bar\theta^2+\sin^2\bar\theta
d\phi^2)]\,,
\eeq
($q=\mu/(2\sqrt{\mu+a^2})$) \ie, the distortion has disappeared and the
solution takes precisely the form of the core of the dilatonic
five-dimensional string. Observe that, for the case $N=3$, the surface
$\rho=0$ is a regular horizon, in fact in this case the geometry is an
$AdS_3\times S^2$ throat, where the coordinate $\psi$ is periodic.
\footnote{Our analysis is a local one. For related global issues, see
\cite{wrapbr}.} The horizon has zero area.

An interesting possibility would be to add momentum running along the
loop. For straight strings this is achieved by boosting the solution
along the direction of the string, but this cannot be done in the
present case. If such a solution were constructed, one could envisage
balancing the system without the need of an external field: the
centrifugal force caused by the rotation of the string might compensate
for the tendency of the loop to collapse. Also, the area of the horizon
might be nonvanishing for this configuration.

Finally, we want to mention that despite the ease with which we have
generalized to arbitrary dilaton coupling the solutions for string
loops, it appears much more difficult to do the same for the solutions
for spherical $p$-branes with $p\geq 2$. Also, non-extremal string
loops (with non-degenerate horizons) would be of interest. For the case
in one dimension less, \ie the dihole solutions, it has been found that
the complication of the metrics increases enormously when considering
the non-extremal versions \cite{nonext}, so perhaps the non-extremal
string loops will be equally complicated.

\section{M5-branes intersecting over a string loop}\label{threem}

Having described how to build string loops for arbitrary $N$ (\ie
arbitrary dilaton coupling $\alpha$), we now turn to see how the
solutions for integer values of $N$, namely $N=1,2,3$ can be obtained
at the intersection of $N$ branes. In particular, we will consider the
intersection of three M5-branes over a string \cite{Ts}. 

In these solutions, the charges of each of the M5-branes will be
independent parameters. Hence, the solution will have three independent
gauge field components. It was already found in \cite{cet} that, in four
dimensions, the step from the dilatonic dihole solutions to the
multicomponent solutions is not too complicated. It works as well here.
We have found the solution for three M5-branes intersecting over a
string loop explicitly as the metric
\beqa
ds^2&=&(T_1 T_2 T_3)^{1/3}\left[-
dt^2+
r^2\cos^2\theta d\psi^2
+{\Si_1\Si_2\Si_3\over
(\De+\gamma^2\sin^2\theta)^2}
\left({dr^2\over\De}+d\theta^2\right)\right]
+{\De\sin^2\theta\over(T_1T_2T_3)^{2/3}} d\varphi^2\nonumber\\
&+&{(T_2T_3)^{1/3}\over T_1^{2/3}}(dy_1^2+dy_2^2)
+{(T_1T_3)^{1/3}\over T_2^{2/3}}(dy_3^2+dy_4^2)
+{(T_1T_2)^{1/3}\over T_3^{2/3}}(dy_5^2+dy_6^2)\,,
\labell{threem5}
\eeqa
and four-form field strength,
\beq
F_{[4]}=3(dA_{(1)}\wedge dy_1\wedge dy_2+dA_{(2)}\wedge dy_3\wedge
dy_4+dA_{(3)}\wedge dy_5\wedge dy_6)\,,
\eeq
with
\beq
A_{(i)}=-{\mu_i a_i\sin^2\theta \over
\De+a_i^2\sin^2\theta}d\varphi\,.
\eeq
The solution depends on four parameters, which correspond to the three
charges of the M5-branes, and the size of the string loop where they
intersect. We have chosen the parameters to be $a_i$ ($i=1,2,3$) and
$\gamma$. Other, non-independent parameters, are defined as
$\mu_i=\gamma^2-a_i^2$. In terms of these, the functions above are
\beqa
\Delta&=&r^2-\gamma^2\,,\nonumber\\
\Sigma_i&=&r^2-a_i^2\cos^2\theta\,,\nonumber\\
T_i&=&{\De+a_i^2\sin^2\theta\over\Si_i}\,.
\eeqa
In the case that the three $a_i$ are all equal, one recovers the
solution \reef{monostring} for $N=3$. If two (one) $a_i$ are equal, and
the other one (two) vanishes, then the solutions for $N=2$ ($N=1$) are
recovered.

All three M5-branes share the direction $\psi$, \ie the string loop.
They intersect pairwise over a tubular three-brane,
$\mathbf{R}^{1,2}\times S^1$, with spatial coordinates $y_i,y_j,\psi$.

As in the cases previously studied, the solution needs to be balanced
by a fluxbrane. Fluxbrane backgrounds can be introduced here by means
of a straightforward generalization of the Harrison transformation for
multiple gauge fields developed in \cite{compo}. The situation is
similar to the one studied in \cite{cet} for quadruple intersections of
branes on diholes, so we will only recapitulate the main features. One
can turn on three different fluxbranes, each along the direction of
either of the three M5-branes. However, it is not necessary to turn on
all three fluxbranes in order to balance the system. Instead, the
conical singularity in the metric can be removed by turning on a single
fluxbrane, which pulls on only one of the M5-branes. This single force
can be sufficient to equilibrate the attraction. When the system is
equilibrated in this fashion, the geometry of the string loop at the
intersection is completely non-singular. However, a certain amount of
deformation remains, in the sense that even if the geometry
approximates $AdS_3\times S^2$, the curvature of each of the two pieces
is not constant. Only when all three intersecting fluxbranes are turned
on, and tuned separately, does the distortion of the horizon disappear
completely. For more details, we refer the reader to the detailed study
in \cite{cet}.

\section{Blown-up strings and branes}\label{blow}

The spherical and tubular branes we have been considering this far carry
no net charge. In this sense, they are the higher dimensional analogues
of electron-positron dipoles. A D-brane, however, can carry along its
worldvolume a gauge field that corresponds to a lower-dimensional brane
that is dissolved in it. Hence, one could envisage situations where the
brane spheres and tubes of the previous sections carry a net charge,
corresponding to some brane that is dissolved in them.

One such situation was considered already some time ago in
\cite{stritun}. It is known that an F-string can be dissolved in the
worldvolume of a D-brane, in the form of an electric worldvolume field.
Hence, it was pointed out in \cite{stritun} that the F-string could
alternatively be viewed as a collapsed Dp-brane tube
$\mathbf{R}^{1,1}\times S^{p-1}$, the Dp-brane carrying an electric
field along the straight brane direction, and the remaining $p-1$
directions being a sphere collapsed to zero size. It was further
observed that this collapse could be prevented by applying a uniform RR
$p+2$-form field strength, to which the Dp-brane couples. The collapsed
brane (\ie the F-string) would be blown up into the
$\mathbf{R}^{1,1}\times S^{p-1}$ tubular brane, and an equilibrium
configuration should be possible. The case of $p=2$ was worked out in
detail in \cite{stritun}, where it was also pointed out that a similar
effect would be expected for higher $p$, as well as for the cases in
which a D-brane can be dissolved inside a higher D-brane. It must be
pointed out that the configurations envisaged in \cite{stritun} were in
unstable equilibrium.

Actually, such configurations of D$p$-branes blown up into
D$(p+2n)$-branes in the presence of a uniform $p+2n+2$-form field
strength were studied in greater detail, and from other perspectives,
in \cite{robs}, and it was remarked that stable configurations would
also be possible. In analogy with electric dipoles, the D-branes behave
as dielectrics. The case of a D0-brane blown up into a spherical
D2-brane is perhaps the simplest, paradigmatic instance of such a
phenomenon.

More recently, a mixture between these two possibilities, with
remarkable features of itself, has been devised in \cite{stubes}. The
system considered consists of a D2-brane tube which carries along its
worldvolume crossed electric and magnetic fields. These correspond to
F-strings and D0-branes, respectively, dissolved in the tube. It turns
out that the angular momentum of this electromagnetic field can balance
the tube tension at a finite radius, in a stable configuration, without
the need of an applied external field. Moreover, such a tube has been
shown to be supersymmetric.

What is missing in all these studies is the introduction of consistent
self-gravity into the systems. It must be mentioned that approximate
solutions for D3-branes blowing up into D5-branes were built in
\cite{polstr}. Our aim here is to include self-gravity in an exact
fashion.

Our approach will be based on the following reasoning. The starting
point in the construction of tubular Dp-branes was the Euclidean
continuation of the higher-dimensional rotating black hole. We are now
interested in tubular branes which carry a net charge, under a field
different to the one they minimally couple to. Since all brane charges
can be generated, via dimensional reduction and then via dualities, from
the charges of the M5 and M2 in eleven dimensions, it appears that the
configurations we seek should be obtainable from the Euclidean
continuation of the solutions for the rotating M5 and M2 brane, and
possibly their intersections too. Fortunately, such solutions have been
constructed in \cite{cvyorot}. We will see that the expectation that
such a procedure may work is borne out in practice.\footnote{A
construction along similar lines was considered in \cite{janmuk}, but
the interpretation of the solutions is missing there.}

In this section we shall describe configurations corresponding to the
three sorts of situations described above: (a) an F-string blown up
into a tubular Dp-brane. (b) a D0-brane blown up into a spherical
D2-brane. (c) an F-string and a D0-brane blown up into a tubular
D2-brane. For the same reasons as explained in Section \reef{prewash},
these configurations, when considered in ten dimensions, will only be
localized along four of the directions transverse to the tubular or
spherical brane. Hence, generically, the branes will be delocalized
along a number of directions.

\subsection{The F1 blown up into a tubular Dp}\label{f1dp}

For definiteness, we will start with the situation where the F-string
blows up into a D6-brane of worldvolume $\mathbf{R}^{1,1}\times S^5$.
To construct this configuration, we start with a solution in eleven
dimension describing a rotating M2-brane \cite{cvyorot}. After 
Wick-rotating the time coordinate, as well as one of the brane
coordinates, one has
\beqa
ds^2_{11}&=&H^{-2/3}\left[-dt^2+dz^2+{\Dea\over \Si}\left(dx_{11}-{\mu
a\cosh\delta\sin^2\theta\over r^4(\Dea)}d\varphi\right)^2\right]
\nonumber\\
&+&H^{1/3}\left[\Si\left({dr^2\over\De}+d\theta^2\right)+
{\Si\De\sin^2\theta\over \Dea} d\varphi^2+r^2\cos^2\theta 
d\Omega^2_5\right]\,,
\eeqa
where
\beqa
\De&=&r^2-a^2-{\mu\over r^4}\,,\nonumber\\
H&=&1+{\mu\sinh^2\delta\over r^4\Si}\,,
\eeqa
and the three-form potentials
\beqa
B_{tzx_{11}}&=&{\mu\cosh\delta\;\sinh\delta\over r^4\Si H},\nonumber\\
B_{tz\varphi}&=&-{a\mu\sinh\delta\;\sin^2\theta\over r^4\Si H}\,.
\labell{threeb}\eeqa
The M2-brane is extended along the directions $z$ and $x_{11}$. In the
situation where the brane is actually rotating, $x_{11}$ would be the
time direction, while $t$ would be a spatial direction along the brane,
but we have Wick-rotated both of them, as well as the rotation
parameter $a$. In this way we are in a situation similar to that in
Section \ref{prewash}, but now a net M2-brane charge is present. The
parameter $\delta$ is associated to this charge. In the limit where
$\mu\to 0$ and $\delta\to\infty$, keeping $\mu e^{2\delta}$ finite, one
recovers the solution for an M2-brane, in prolate spheroidal
coordinates.

The solution can now be reduced down to ten dimensions along the
direction of $x_{11}$, to a solution of type IIA string theory.  Notice
that the string coupling constant, which is determined by the asymptotic
radius of the compact circle, is fixed by the periodicity requirements
for $x_{11}$.
In the
process, the M2-brane becomes a F-string along the $z$ direction. But,
as expected from the analysis in the previous sections, we also
generate a tubular D6-brane, with worldvolume geometry ${\bf
R}^{1,1}\times S^5$. Explicitly, the string metric for the solution is
\beqa
ds^2=&&\left({\Dea\over\Si}\right)^{1/2}\biggl[H^{-1}(-dt^2+dz^2)+
\Si\left({dr^2\over\De}+d\theta^2\right)\nonumber\\
&&+r^2\cos^2\theta d\Omega^2_5+
{\Si\De\sin^2\theta\over \Dea} d\varphi^2\biggr]\,,
\eeqa
with dilaton 
\beq
e^\phi=H^{-1/2}\left({\Dea\over\Si}\right)^{3/4}\,,
\eeq
and potentials
\beqa
A_\varphi&=&-{\mu a\cosh\delta\;\sin^2\theta\over 
r^4(\Dea)}\,,\nonumber\\
B_{tz}&=&{\mu\cosh\delta\;\sinh\delta\over r^4\Si H}\,,\nonumber\\
B_{tz\varphi}&=&-{a\mu\sinh\delta\;\sin^2\theta\over r^4\Si H}\,.
\eeqa
$A_\varphi$ is created by the D6-brane tube, while $B_{tz}$ is the
Kalb-Ramond potential for the F-string. The remaining $B_{tz\varphi}$
appears as a consequence of the mixing between the latter two in the
equations of motion, and vanishes whenever only one of the two sorts of
charges is present.

One may question whether the F-string has expanded along with the
D6-brane, or instead it remains back at the center of the sphere $S^5$.
The correct answer is actually the former. First, note that if the
F-string were inside the sphere, the core of the string should appear
as a singularity at $r=r_+$, and at some $\theta\in[0,\pi/2]$ (recall
that $\theta=\pi/2$ corresponds to the center of the sphere, and
$\theta=0$ to the boundary). But, as a matter of fact, at $r=r_+$, and
any value of $\theta$, the function $H$ remains finite. It follows from
a trivial extension of our analysis in Section \ref{prewash} that, in
this case, the only singularity present inside the sphere is the
expected conical defect.

The evidence that the F-string is actually dissolved in the D6-brane
tube, at $r=r_+$ and $\theta=0$, is borne out from the study of the
solution near the tube. Performing the change of coordinates
\reef{transform}, and going near the core (\ie to small $\rho$),
results into $H\to\cosh^2\delta$ and
\beqa
ds^2&=&\left({g(\bar\theta)\rho\over
q}\right)^{1/2}\left[{-dt^2+dz^2\over\cosh^2\delta}+r_+^2d\Omega^2_5+
{q\over \rho}(d\rho^2+\rho^2d\bar\theta^2)\right]\nonumber\\
&+&\left({q \over g(\bar\theta)\rho}\right)^{1/2}\rho^2\sin^2\bar\theta
d\varphi^2\,,
\labell{nearfd6}\eeqa
where $g(\bar\theta)$ is as in \reef{gtheta}, and $q=(r_+^2-a^2)/\Delta'_+$
is the same as was defined after \reef{nearmet}.

This should reproduce the solution for an F-string dissolved into a
flat D6-brane, up to the angular distortion factors, which are due to
unbalanced forces, and the replacement of the 5-sphere by a flat
$\mathbf{R}^5$. The string-frame metric for such a bound state was
constructed in \cite{copa}, as
\beq
ds^2=\tilde f^{1/2}[f^{-1}(-dt^2+dz^2)+\tilde f^{-
1}d\mathbf{x}^2_{(5)}+d\rho^2+\rho^2d\Omega_2^2]\,,
\eeq
with
\beq
f=1+{k\over\rho}\,,\qquad
\tilde f=1+{k\sin^2\alpha\over\rho}\,.
\eeq
The D6-brane is recovered for $\alpha=\pi/2$, and the delocalized 
F-string appears for $\alpha=0$.
The core limit is $\rho\to 0$, in which the metric becomes
\beq
ds^2=\left({\rho\over k\sin^2\alpha}\right)^{1/2}[\sin^2\alpha(-
dt^2+dz^2)+d\mathbf{x}^2_{(5)}]+\left({\rho\over
k\sin^2\alpha}\right)^{-1/2}(d\rho^2+\rho^2d\Omega_2^2)\,.
\labell{f1d6core}\eeq
Notice that the singular behavior near the core is dominated by the
D6-brane. Comparing to \reef{nearfd6}, we can identify the solutions, up
to the distortion factors, by making $\cosh\delta=1/\sin\alpha$, and
$q=k\sin^2\alpha$. Although we have not written them explicitly, one can
easily check that the one-, two-, and three-form potentials also map
correctly between both configurations, up to the distortion introduced
by $g(\bar\theta)$.

The way to proceed now should be clear from what we have seen in the
previous sections. A RR fluxbrane background that couples to the
D6-brane can be introduced, via a twist in the reduction from eleven to
ten dimensions. One then finds the string-frame metric
\beq
ds^2=\Lambda^{1/2}\left[H^{-1}(-dt^2+dz^2)+
\Si\left({dr^2\over\De}+d\theta^2\right)
+r^2\cos^2\theta d\Omega^2_5+
{\De\sin^2\theta\over \Lambda} d\varphi^2\right]\,,
\eeq
and dilaton $e^\phi=H^{-1/2}\Lambda^{3/4}$, with
\beqa
\Lambda&=&{\Dea+2B\mu r^{-4}
a\cosh\delta\sin^2\theta+B^2\sin^2\theta[(r^2-a^2)^2+\De
a^2\sin^2\theta]\over\Si}\nonumber\\
&+&{B^2a^2\mu^2r^{-4}\sinh^2\delta\sin^2\theta\over \Si(\Dea)}\,.
\labell{lambdel}\eeqa

This fluxbrane exerts a pressure on the D6-brane tube, and the strength
of the fluxbrane can be tuned so as to balance the system, \ie cancel
the conical singularities at $r=r_+$. This happens for 
\beq
B={6r_++4a\over 2r_+(r_++a)\cosh\delta}
\eeq
(c.f.
eq.~\reef{chooseb}). For this value of the fluxbrane
strength, the angular distortion of the core disappears,
$g(\bar\theta)=1$, and we
recover precisely \reef{f1d6core}, with $r_+^2d\Omega_5^2$ instead of
$d\mathbf{x}^2_{(5)}$.

In order to build F-strings blown up into other tubular Dp-branes, with
worldvolume $\mathbf{R}^{1,1}\times S^{p-1}$ (and delocalized along $6-
p$ directions), consider the eleven-dimensional metric
\beqa
ds^2_{11}&=&H^{-2/3}\left[-dt^2+dz^2+{\Dea\over \Si}\left(dx_{11}-{\mu
a\cosh\delta\sin^2\theta \over r^{p-2}(\Dea)}d\varphi\right)^2\right]
\nonumber\\
&+&H^{1/3}\left[d\mathbf{x}^2_{(6-
p)}+\Si\left({dr^2\over\De}+d\theta^2\right)+
{\Si\De\sin^2\theta\over \Dea} d\varphi^2+r^2\cos^2\theta
d\Omega^2_{p-1}\right]\,,
\eeqa
where now $\De=r^2-a^2-\mu r^{2-p}$, 
$H=1+{\mu\sinh^2\delta\over r^{p-2}\Si}$,
and the three-form potentials are as in \reef{threeb}, but with
$r^{p-2}$ in the denominators, instead of $r^4$.

In a manner entirely analogous to the one just discussed, when reduced
down to a solution of ten-dimensional type IIA supergravity, this
results into an F-string spread (dissolved) along the worldvolume
$\mathbf{R}^{1,7-p}\times S^{p-1}$ of a D6-brane. Then, T-duality along
the straight $7-p$ directions yields the desired solution. The way to
proceed is straightforward, and therefore we shall not give any further
details.

\subsection{The D0 blown up into a D2 sphere}

The construction is very similar to the one in the previous subsection,
but this time one starts with the metric for a rotating M5-brane
\cite{cvyorot}\footnote{A few typos in \cite{cvyorot} were corrected in
\cite{corr}.},
Wick-rotated along the time direction and one spatial brane direction:
\beqa
ds^2_{11}&=&H^{-1/3}\left[-dt^2+d\mathbf{x}_{(4)}^2+{\Dea\over
\Si}\left(dx_{11}-{\mu
a\cosh\delta\sin^2\theta\over r(\Dea)}d\varphi\right)^2\right]
\nonumber\\
&+&H^{2/3}\left[\Si\left({dr^2\over\De}+d\theta^2\right)+
{\Si\De\sin^2\theta\over \Dea} d\varphi^2+
r^2\cos^2\theta d\Omega^2_2\right]\,,
\eeqa
with $\De=r^2-a^2-\mu r^{-1}$ and $H=1+{\mu\sinh^2\delta\over r\Si}$.
Reduction along $x_{11}$ results into a D4-brane that has grown a
D6-brane with worldvolume $\mathbf{R}^{1,4}\times S^2$. By twisting the
direction of the reduction, as described in Section \ref{prewash}, the
conical singularities at $r=r_+$ can be cancelled out and the system be
equilibrated.

In order to obtain the configuration for a D0-brane blown up into the
D2-brane sphere, one simply performs a T-duality transformation along
the directions of the D4. As a result, the D0-D2 configuration is
delocalized in these four directions. The resulting string-frame metric
is
\beqa
ds^2&=&-H^{-1/2}\Lambda^{1/2}dt^2+H^{1/2}\Lambda^{1/2}\left[
\Si\left({dr^2\over\De}+d\theta^2\right)
+r^2\cos^2\theta d\Omega^2_2+
{\De\sin^2\theta\over \Lambda}
d\varphi^2\right]\nonumber\\
&+&H^{1/2}\Lambda^{-1/2}d\mathbf{x}_{(4)}^2\,,
\eeqa
and dilaton $e^\phi=H^{3/4}\Lambda^{-1/4}$. The function $\Lambda$ is
as in \reef{lambdel}, but now where $r^{-4}$ appears, it must read
$r^{-1}$. The system is equilibrated for
\beq
B={3r_++a\over 2r_+(r_++a)\cosh\delta}\,.
\eeq

\subsection{The F1 and D0 in the D2 tube}

This configuration has the same structure as the ``supertube''
discussed in \cite{stubes} (see also \cite{bak}). Recall that, in the
Dirac-Born-Infeld picture, such a supertube could be held in stable
equilibrium at a finite radius of the tube, without the need of a RR
external field to support the D2 tube against collapse.

The eleven-dimensional starting point is the rotating
intersecting M2-M5 system. These intersect over a string. By
Wick-rotating both the time and the space directions along this string,
as well as the rotation parameter $a$, one finds
\beqa
ds^2_{11}&=&H_m^{-1/3}H_e^{-2/3} \left[-dt^2+{\Dea\over
\Si}\left(dx_{11}-{\mu
a\cosh\delta_e\cosh\delta_m 
\sin^2\theta \over\Dea}d\varphi\right)^2\right]
\nonumber\\
&+& H_m^{2/3}H_e^{-2/3}dz^2+H_m^{-1/3}H_e^{1/3}d\mathbf{x}^2_{(4)}
\nonumber\\
&+&
H_m^{2/3}H_e^{1/3}\left[\Si\left({dr^2\over\De}+d\theta^2\right)+
{\Si\De\sin^2\theta\over \Dea} d\varphi^2+r^2\cos^2\theta
d\psi\right]\,,
\eeqa
where now $\De=r^2-a^2-\mu$, and
\beq
H_{e,m}=1+{\mu\sinh^2\delta_{e,m}\over \Si}\,.
\eeq
The expressions for the different components of the three-form potential
can be obtained from \cite{cvyorot}. We shall not need them here, though.

The indices $e$, $m$, refer to the electric M2 and magnetic M5,
respectively. The M2 and M5 share the direction $x_{11}$, which is the
one to be wrapped. In addition to this, the M2 extends also along $z$,
while the M5 spans the four coordinates $\mathbf{x}_{(4)}$. Hence,
reduction to type IIA theory will result into an F1 extended along $z$,
and a D4 along $\mathbf{x}_{(4)}$. Both will be dissolved into a tubular D6
which spans $(z,\mathbf{x}_{(4)},\psi)$. In order to achieve the
configuration we seek, we T-dualize the four D4 directions. The metric,
in string frame, is
\beqa
ds^2&=&\left({\Dea\over
\Si}\right)^{1/2}\Biggl\{-H_m^{-1/2}H_e^{-1} dt^2+
H_m^{1/2}H_e^{-1}dz^2+H_m^{-1/2}d\mathbf{x}^2_{(4)}
\nonumber\\
&+&
H_m^{1/2}\left[\Si\left({dr^2\over\De}+d\theta^2\right)+
{\Si\De\sin^2\theta\over \Dea} d\varphi^2+r^2\cos^2\theta
d\psi\right]\Biggr\}\,.
\eeqa

An analysis of the geometry near the core of the D2 tube, along the
lines of that in subsection \ref{f1dp}, shows that near this core the
solution reproduces the core of the bound state of an F1 and a D0
dissolved in the worldvolume of a flat D2.

Now we address the question of whether this configuration might be
balanced in the absence of a fluxbrane. To this effect, we analyze the
possible conical singularities on the fixed-point set of the vector
$\partial_\varphi$, \ie the disk at $r=r_+=\sqrt{\mu+a^2}$. Since a
conformal rescaling of the
metric coefficients $g_{rr}$ and $g_{\varphi\varphi}$ does not affect
the conical structure, it is easy
to see that the result is the same as \reef{cone}, namely,
\beq
\lim_{r\to r_+}{2\pi\over\sqrt{g_{rr}}}
{d\sqrt{g_{\varphi\varphi}}\over dr}={2\pi\sqrt{\mu+a^2}\over
a}\,,
\eeq
which is independent of the D0 and F1 charges, and does not equal $2\pi$
for any non-trivial choice of parameters. The conical singularity
remains. Disappointingly, the tube, by itself, cannot be in equilibrium
(stable or unstable), whatever the value of the D0 and F1 charges.

\section{Discussion}

The conclusion of the analysis of this last configuration is puzzling,
and it poses the following question: To what extent do the
configurations we have obtained correspond to similar configurations,
which have been studied earlier in the absence of self-gravity (closed
string effects)?

Let us specify the terms for the comparison. A convenient way to study
brane tubes and spheres, from the point of view of open strings, is by
using the Dirac-Born-Infeld description of the worldvolume dynamics of
a D-brane. In order to find the equilibrium configurations of these
systems, one studies their static potential energy, for a given field
strength and brane charge, as a function of the radius of the tube (or
sphere). For an F-string of charge $q_s$, blown up into a tubular
D$p$-brane under the action of a RR $p+2$-form field of strength $B$,
this potential was computed in \cite{stritun}, with the result
\beq
V_p(R)= \sqrt{R^{2p-2}+q_s^2}-{B\over p}R^{p}\,,
\labell{pbrv}
\eeq
in appropriate units. Tubes in equilibrium under the action of the
external field have radii that correspond to the extrema of this
potential. For $p=2$, the shape of this potential is the following:
$R=0$ is a stable minimum, and if $B<|q_s|^{-1}$, there is a maximum at
$R=\sqrt{B^{-2}-q_s^2}$. By contrast, for $p\geq 3$, the extremum at
$R=0$ becomes a local unstable maximum, and a minimum appears at a
finite value of the radius $R$. For larger values of $R$ (and $B$ less
than a critical value) one finds again a (global) maximum, with a large
value of $V_p$, \ie an unstable equilibrium configuration.

Essentially the same potential as \reef{pbrv} describes a D$q$-brane
dissolved in a tubular D$(q+2n)$-brane, after identifying $2n=p-1$
\cite{stritun}. The minimum of this potential for the case of a D0
expanded into a spherical D2 (\ie the case $p=3$ for \reef{pbrv})
corresponds to a dielectric configuration. These were studied in
\cite{robs}, where the existence of such configurations was also
established from the point of view of matrix theory.

On the other hand, for the supertube, the potential energy, in the
absence of any external field, is
\cite{stubes}
\beq
V(R)={1\over R}\sqrt{(q_s^2+R^2)(q_0^2+R^2)}\,,
\labell{stubev}\eeq
where $q_s$ and $q_0$ are the F-string and D0-brane charges,
respectively. This potential possesses a unique
extremum, a minimum at $R=\sqrt{|q_sq_0|}$.

Obviously, we have been unable to find, in our supergravity
configurations, this minimum of the supertube potential. No such stable
equilibrium configuration appears to be possible. This leads us to
question whether the other configurations, namely the F1 in the Dp
tube, and the D0 in the D2 sphere, can correspond to the configurations
at the minima of the potentials. In this respect, notice that the
Dirac-Born-Infeld potential \reef{pbrv} does not yield any stable
minimum for the radius for the case of an F-string expanded into a D2
tube, so in this case it appears fairly clear that the gravitating
configuration must correspond to this unstable point. Given these
considerations, it may sound plausible that, also for $3\leq p\leq 6$,
the configurations we have built are associated to the tubes at the
unstable maxima. Otherwise, one should expect to find, for $3\leq p\leq
6$, two different equilibrium configurations, one stable and the other
one unstable, for fixed values of the string charge, the external
field, and, possibly, the number of tubular $p$-branes.

Similar considerations apply to the solution for a D0-brane blown up
into a D2 sphere. In his study of dielectric branes, \cite{robs}, Myers
considered a regime in which one zooms in on the region between zero
radius and the stable minimum. In this regime, the unstable maximum is
pushed to a much larger value of the radius, and to a very high energy.
Effectively, it disappears. It may be that the situation described in
this paper is the opposite, in which only the unstable maximum is seen.

These observations, however, are not conclusive, and in order to reach
a firmer verdict, a more detailed study of these configurations would
be necessary. One possibility is to study the energetics of the
supergravity solutions, possibly considering the off-shell
configurations where the conical singularity is not eliminated. Such
off-shell configurations would presumably correspond to points other
than the extrema in the potentials above. Perhaps in this way one can
determine whether the equilibrium configurations sit at a maximum or a
minimum. 

Another possibility, at first sight much more complicated, is to
address directly the classical linearized stability of the
configurations. As observed in \cite{dggh}, the instability of the
tubular $p$-branes (without a net charge) is manifested by the
existence of an unstable mode in the fluctuations around the Euclidean
black holes \cite{gpy}\footnote{This is not to be confused with the
instability of black branes \cite{gl}, whose onset is also associated
to this same mode, as has been explicitly mentioned in \cite{reall}.}.
Perhaps this argument can be modified so as to yield a proof of the
instability of all the configurations described in this paper. If this
were the case, a description of the dielectric effect that accounts for
the self-gravity of the branes would remain an elusive problem.

At any rate, it appears clear that a detailed analysis of the
properties of these solutions is required. In this paper we have
applied and extended the techniques of \cite{dggh}, and managed to
construct a variety of solutions describing exact tubular branes in
string and M-theory. Our focus has been mostly on the construction of
the solutions. Further study of their properties is left for future
work.

\section*{Acknowledgments}

Partial support from UPV grant 063.310-EB187/98 and CICYT AEN99-0315 is
acknowledged.

\end{document}